\documentclass[12pt,a4paper]{article}
\usepackage[doublespacing]{setspace}
\usepackage[latin1]{inputenc}
\usepackage{amsmath}
\usepackage{amsfonts}
\usepackage{amssymb}
\usepackage{hyperref}

\usepackage[english]{babel}
\usepackage{amsmath,amsthm}
\usepackage{amsfonts}
\usepackage{appendix}
\usepackage{rotating}

\usepackage{algorithm}
\floatname{algorithm}{Algorithm}

\usepackage{natbib}
\usepackage[all,cmtip]{xy}

\newtheorem{prop}{Proposition}

\begin{document}

\title{Inference for two-stage sampling designs with application to a panel for urban policy}

\author{Guillaume Chauvet$^{(1)}$ and Audrey-Anne Vall\'ee$^{(2)}$ \\
$^{(1)}$ Ensai (Irmar), Campus de Ker Lann, Bruz - France \\
$^{(2)}$ Institut de Statistique, Universit\'e de Neuch\^atel, Switzerland}

\maketitle

\begin{abstract}
\noindent Two-stage sampling designs are commonly used for household and health surveys. To produce reliable estimators with assorted confidence intervals, some basic statistical properties like consistency and asymptotic normality of the Horvitz-Thompson estimator are desirable, along with the consistency of assorted variance estimators. These properties have been mainly studied for single-stage sampling designs. In this work, we prove the consistency of the Horvitz-Thompson estimator and of associated variance estimators for a general class of two-stage sampling designs, under mild assumptions. We also study two-stage sampling with a large entropy sampling design at the first stage, and prove that the Horvitz-Thompson estimator is asymptotically normally distributed through a coupling argument. When the first-stage sampling fraction is negligible, simplified variance estimators which do not require estimating the variance within the Primary Sampling Units are proposed, and shown to be consistent. An application to a panel for urban policy, which is the initial motivation for this work, is also presented.
\end{abstract}


\section{Introduction} \label{sec1}

\noindent In household and health surveys, the population is often sparse over a large territory and there is regularly no sampling frame. Two-stage sampling designs are convenient in such situations. The population are grouped into large blocks (e.g., municipalities or counties), called Primary Sampling Units (PSUs), which are sampled at the first stage. Only a frame of these PSUs is needed at this stage, which is easier to create. At the second stage, a list of population units is obtained inside the selected PSUs, and a sample of these population units is selected. Despite its convenience, multistage sampling has the drawback to lead to estimators with inflated variance, compared to sampling designs where the population units are directly selected. A detailed treatment of multistage sampling may be found in \cite{coc:77}, \cite{sar:swe:wre:92} and \cite{ful:11}. \\

\noindent To produce reliable estimators with assorted confidence intervals, some statistical properties are needed for a sampling design: (a) the Horvitz-Thompson estimator should be consistent for the true total; also, (b) this estimator should be asymptotically normally distributed, and (c) consistent variance estimators should be available, to be able to produce normality-based confidence intervals. General conditions for the consistency of the Horvitz-Thompson estimator are given in \cite{isa:ful:82} and \cite{rob:82}, see also \cite{pra:sen:09}. The asymptotic normality is usually studied design by design, see for example \cite{haj:64} for rejective sampling, \cite{ros:72} for successive sampling or \cite{ohl:86} for the Rao-Hartley-Cochran~(1972) \nocite{rao:har:coc:62} procedure; see also \cite{bic:fre:84} for stratified simple random sampling and \cite{che:rao:07} for two-phase sampling designs.
These properties are also studied in \cite{bre:ops:00} for the class of local polynomial regression estimators and in \cite{bre:ops:san:16}, but under assumptions that are not generally applicable for multistage sampling designs. More recently, \cite{boi:lop:rui:17} and \cite{ber:cha:cle:17} established functional central limit theorems for Horvitz-Thompson empirical processes. In summary, these properties have been mainly studied in the literature for one-stage sampling designs. \\

\noindent In two-stage sampling, the asymptotic properties of estimators are more difficult to study, due to the dependence introduced in the selection of the sampling units.
\cite{kre:rao:81} studied the case when the primary units are selected with replacement, and \cite{ohl:89} derived a general central limit theorem for such designs.
Recently, \cite{cha:15} considered coupling methods to prove the asymptotic normality of the Horvitz-Thompson estimator and the validity of a bootstrap procedure for stratified simple random sampling at the first stage.
However, there is a lack of general conditions ensuring that properties (a)-(c) hold for general two-stage sampling designs, and this is the purpose of the present paper. A notable exception is \cite{bre:ops:08}, who obtain the consistency of the Horvitz-Thompson estimator under very weak conditions. This is discussed in Section \ref{sec4}. \\

\noindent In this paper, the properties of estimators and variance estimators are studied for a general class of two-stage sampling designs.
The framework is introduced in Section \ref{sec2} and the variance of the Horvitz-Thompson estimator is decomposed in a sum of three components. In Section \ref{sec3}, the assumptions used to establish the asymptotic properties are defined.
In Section \ref{sec4}, the Horvitz-Thompson estimator is shown to be consistent under our conditions, and the order of magnitude of the three components of the variance is determined. The consistency of two unbiased variance estimators is established in Section \ref{sec41}.
A simplified variance estimator which does not require estimating the variance within the PSUs can be produced.
We prove in Section \ref{sec42} that this variance estimator is consistent when the total variance within the PSUs is negligible. In Section \ref{sec5}, the specific case of large-entropy sampling designs at the first-stage is considered. When rejective sampling is used at the first-stage, the consistency of a H\'ajek-type variance estimator is established under reduced assumptions, along with the asymptotic normality of the Horvitz-Thompson estimator. We define a coupling procedure to extend these results to a more general class of large-entropy sampling designs at the first-stage. In Section \ref{sec6}, the properties of the H\'ajek-type variance estimators are evaluated in a simulation study. An application to a panel for urban policy, which is the initial motivation for this work, is presented in Section \ref{appli:ppv}.


\section{Notation} \label{sec2}

\noindent We are interested in a finite population $U$ of size $N$, in which a sample is selected by means of a two-stage sampling design. The units in $U$, called Secondary Sampling Units (SSUs) are partitioned into a population $U_I$ of $N_I$ Primary Sampling Units (PSUs). A sample $S_I$ of $n_I$ PSUs is selected in $U_I$. We are interested in estimating the population total
    \begin{eqnarray} \label{eq1:sec2}
    Y & = & \sum_{i=1}^{N_I} \sum_{k =1}^{N_i} y_{ik}
    =\sum_{i=1}^{N_I} Y_i,
    \end{eqnarray}
for some variable of interest $y$, where $Y_i=\sum_{k =1}^{N_i} y_{ik}$ is the sub-total of the variable $y$ on the PSU $i$ and $N_i$ is the number of SSUs inside the PSU $i$. \\

\noindent We assume that the population $U$ belongs to a nested sequence $\{U_t\}$ of finite populations with increasing sizes $N_t$, and that the population vector of values $y_{Ut}=(y_{1t},\ldots,y_{Nt})^{\top}$ belongs to a sequence $\{y_{Ut}\}$ of $N_t$-vectors. The index $t$ is suppressed in what follows but all limiting processes are taken as $t \to \infty$. We assume that $N_I \to \infty$ and $n_I \to \infty$ as $t \to \infty$. We consider a single stratum of PSUs, but our results may be easily generalized to the case of a finite number of strata, see the application to the panel for urban policy in Section \ref{appli:ppv}. An alternative asymptotic set-up is possible, under which the number of strata tends to infinity while the sample size per stratum remains bounded, see \cite{kre:rao:81} and \cite{bre:ops:san:16}. \\

\noindent We note $I_{Ii}$ for the sample membership indicator of the PSU $i$ into $S_I$, $\pi_{Ii}=E(I_{Ii})$ for the inclusion probability of the PSU $i$, and $\pi_{Iij} = E(I_{Ii} I_{Ij})$ for the probability that the PSUs $i$ and $j$ are selected jointly in $S_I$. Inside any PSU $i \in S_I$, a sample $S_i$ of $n_i$ SSUs is selected at the second stage. We note
    \begin{eqnarray} \label{eq2:sec2}
      N_0 = \frac{1}{N_I} \sum_{i=1}^{N_I} N_i & \textrm{ and } & n_0 = \frac{1}{N_I} \sum_{i=1}^{N_I} n_i
    \end{eqnarray}
for the average size of the PSUs and for the average sample size selected inside the PSUs. We do not need particular assumptions on the limit behaviour of $n_0$ and $N_0$, and $n_0$ may be either bounded or unbounded. Our set-up covers in particular the case when the SSUs are comprehensively surveyed inside a selected PSU, which amounts to single-stage sampling on the population of PSUs. \\

\noindent For any SSU $k$ in the PSU $i$, we note $I_{k}$ for the sample membership indicator of $k$ in $S_i$. Also, we note $\pi_{k|i}=E(I_{k}|i \in S_I)$ for the conditional inclusion probability of $k$, and $\pi_{kl|i} = E(I_{k} I_{l}|i \in S_I)$ for the conditional joint probability that two SSUs $k,l \in i$ are selected together in $S_i$. We assume invariance of the second-stage designs, as defined by \cite{sar:swe:wre:92}: the second stage of sampling is independent of $S_I$. Also, we assume that the second-stage designs are independent from one PSU to another, conditionally on $S_I$. \\

\noindent The Horvitz-Thompson (HT) estimator of $Y$ is
    \begin{eqnarray} \label{eq3:sec2}
      \hat{Y}_{\pi} = \sum_{i \in S_I} \frac{\hat{Y}_i}{\pi_{Ii}} & \textrm{ with } & \hat{Y}_i = \sum_{k \in S_i} \frac{y_{ik}}{\pi_{k|i}}.
    \end{eqnarray}
The variance of $\hat{Y}_{\pi}$ may be written as
    \begin{eqnarray} \label{eq4:sec2}
      V(\hat{Y}_{\pi}) & = & \sum_{i=1}^{N_I} \sum_{j=1}^{N_I}  \Delta_{Iij} \frac{Y_i}{\pi_{Ii}} \frac{Y_j}{\pi_{Ij}} + \sum_{i=1}^{N_I}  \left(\frac{1-\pi_{Ii}}{\pi_{Ii}}\right) V_i + \sum_{i=1}^{N_I}  V_i \nonumber \\
     & = & V_1(\hat{Y}_{\pi})+V_2(\hat{Y}_{\pi})+V_3(\hat{Y}_{\pi})
    \end{eqnarray}
with $\Delta_{Iij}=\pi_{Iij}-\pi_{Ii}\pi_{Ij}$, and
    \begin{eqnarray} \label{eq5:sec2}
      V_i \equiv V(\hat{Y}_i) & = & \sum_{k=1}^{N_i}\sum_{l=1}^{N_i} \Delta_{kl|i} \frac{y_{ik}}{\pi_{k|i}} \frac{y_{il}}{\pi_{l|i}},
    \end{eqnarray}
with $\Delta_{kl|i}=\pi_{kl|i}-\pi_{k|i}\pi_{l|i}$. The term $V_1(\hat{Y}_{\pi})$ is the variance due to the first stage. The sum of the two last terms in (\ref{eq4:sec2}) may be simplified as
    \begin{eqnarray} \label{eq6:sec2}
      V_2(\hat{Y}_{\pi})+V_3(\hat{Y}_{\pi}) & = & \sum_{i=1}^{N_I} \frac{V_i}{\pi_{Ii}}.
    \end{eqnarray}
This is the variance due to the second stage of sampling. \\

\noindent When estimating the variance, the terms $V_1(\hat{Y}_{\pi})+V_2(\hat{Y}_{\pi})$ and $V_3(\hat{Y}_{\pi})$ are handled separately. Variance estimators for these two terms are considered in Section \ref{sec4}, and proved to be consistent under assumptions which are stated and discussed in Section \ref{sec3}.
In case of large entropy sampling designs at the first-stage, consistent variance estimators can be produced under reduced assumptions, and without using second-order inclusion probabilities. This is studied in Section \ref{sec5} .

\section{Assumptions} \label{sec3}

\noindent To study the asymptotic properties of the estimators and variance estimators that we consider below, a number of assumptions are needed. We present in Section \ref{sec21} the assumptions on the first-stage sampling design, and in Section \ref{sec22} the assumptions on the second-stage sampling designs. The assumptions related to the variable of interest are presented in Section \ref{sec23}.

\subsection{Assumptions on the first-stage sampling design} \label{sec21}


    \begin{itemize}
      \item[FS1:] Some constant $f_{I0}<1$ exists s.t.
        \begin{eqnarray} \label{eq1:sec3}
          N_I^{-1} n_I & \leq  & f_{I0}.
        \end{eqnarray}
      Some constants $c_{I1},C_{I1}>0$ exist s.t. for any PSU $i$
        \begin{eqnarray} \label{eq2:sec3}
          c_{I1} & \leq N_I n_I^{-1} \pi_{Ii} \leq & C_{I1}.
        \end{eqnarray}
      \item[FS2:] Some constants $C_{I2},C_{I3}>0$ exist s.t. for any PSUs $i \ne j \ne {i'}$
        \begin{eqnarray}
            & \pi_{Iij} \leq & C_{I2} N_I^{-2} n_I^2, \label{eq3:sec3} \\
            & \pi_{Iiji'} \leq & C_{I3} N_I^{-3} n_I^3,  \label{eq4:sec3}
        \end{eqnarray}
      with $\pi_{Iiji'}$ the probability that the PSUs $i,j,{i'}$ are selected together in $S_I$.
      Some constants $C_{I4},C_{I5}$ exist s.t.
        \begin{eqnarray}
          \Delta_{I1} \equiv \max_{i \neq {j}=1,\ldots,N_I} \left| \pi_{Iij}-\pi_{Ii} \pi_{Ij} \right| \leq C_{I4} N_I^{-2} n_I, \label{eq5:sec3} \\
          \Delta_{I2} \equiv \max_{i \neq {j} \neq {i'} \neq {j'} =1,\ldots,N_I} \left| \pi_{Iiji'j'}-\pi_{Ii} \pi_{Ij}\pi_{Ii'} \pi_{Ij'} \right| \leq C_{I5}  N_I^{-4} n_I^3, \nonumber 
        \end{eqnarray}
      with $\pi_{Iiji'j'}$ the probability that the PSUs $i,j,{i'},{j'}$ are selected together in $S_I$.
      \item[FS3:] Some constant $c_{I2}>0$ exists s.t. for any $i \ne j =1,\ldots,N_I$
        \begin{eqnarray} \label{eq6b:sec3}
           c_{I2} N_I^{-2} n_I^2 & \leq \pi_{Iij}.
        \end{eqnarray}
    \end{itemize}

\noindent The Assumption (FS1) is related to the order of magnitude of the first-stage sample size $n_I$, and to the first-order inclusion probabilities. Equation (\ref{eq1:sec3}) ensures that the first-stage sample is not degenerate, in the sense that the PSUs are not comprehensively surveyed. This assumption is compatible with the case $n_I/N_I \to 0$ (negligible first-stage sampling fraction). A similar condition is considered in \cite[][assumption A5]{bre:ops:00}, and in \cite[][assumption HT3]{boi:lop:rui:17}. Equation (\ref{eq2:sec3}) states that the first-order inclusion probabilities do not depart much from that obtained under simple random sampling. The same condition is considered in \cite[][assumption C1]{boi:lop:rui:17}. Overall, (FS1) is under the control of the survey sampler. \\

\noindent The Assumption (FS2) is related to the inclusion probabilities of order $2$ to $4$. If (FS1) holds, equations (\ref{eq3:sec3}) and (\ref{eq4:sec3}) will automatically hold for negatively associated sampling designs \citep[e.g.][]{bra:jon:12} which includes simple random sampling, rejective sampling \citep{haj:64}, Sampford sampling \citep{sam:67} and pivotal sampling \citep{dev:til:98,cha:12}, for example. In equation (\ref{eq5:sec3}), 
the quantities $\Delta_{I1}$ and $\Delta_{I2}$ are two measures of dependency in the selection of units. These quantities will be equal to $0$ when the units are selected independently, which is known as Poisson sampling \citep[see for example][p. 13]{ful:11}. Equation (\ref{eq5:sec3}) is respected for simple random sampling. If (FS1) holds, it is also respected under rejective sampling \citep[see][Theorem 1]{boi:lop:rui:12}, and it can be proved that it holds for the Rao-Sampford sampling design (see Hajek, 1981, Chapter 8). Similar conditions are considered in \cite[][assumption A7]{bre:ops:00}, and in \cite[][conditions C2-C4]{boi:lop:rui:17}. \\

\noindent The Assumption (FS3) provides a uniform lower bound for the second-order inclusion probabilities. A similar condition is considered in \cite[][assumption A6]{bre:ops:00}. This assumption holds for simple random sampling, but is more difficult to prove for unequal probability sampling designs. On the other hand, it is needed to prove the consistency of the Horvitz-Thompson variance estimator and of the Yates-Grundy variance estimator, see our Section \ref{sec41} and Theorem 3 in \cite{bre:ops:00}. We consider in Section \ref{sec5} the specific case of large entropy sampling designs at the first-stage, for which alternative consistent variance estimators are possible, and for which the assumption (FS3) can be suppressed.

\subsection{Assumptions on the second-stage sampling design} \label{sec22}

    \begin{itemize}
      \item[SS0:] Some constants $\lambda_1,\Lambda_1>0$ and $\phi_1,\Phi_1>0$ exist s.t. for any PSU $i$
        \begin{eqnarray}
          \lambda_1 n_0 & \leq n_i \leq & \Lambda_1 n_0, \label{eq8:sec3} \\
          \phi_1 N_0 & \leq N_i \leq & \Phi_1 N_0. \label{eq9:sec3}
        \end{eqnarray}
      \item[SS1:] Some constants $c_{1},C_{1}>0$ exist s.t. for any PSU $i$ and for any $k$ inside:
        \begin{eqnarray} \label{eq10:sec3}
          c_{1} & \leq N_0 n_0^{-1} \pi_{k|i} \leq & C_{1}.
        \end{eqnarray}
      \item[SS2:] Some constants $C_2,C_3>0$ exists s.t. for any PSU $i$ and any $k \ne l \ne k'$ inside:
        \begin{eqnarray}
           & \pi_{kl|i} \leq & C_{2} N_0^{-2} n_0^2, \label{eq11:sec3} \\
           & \pi_{klk'|i} \leq & C_{3} N_0^{-3} n_0^3, \label{eq12:sec3}
        \end{eqnarray}
      with $\pi_{klk'|i}$ the conditional probability that the SSUs $k,l,k'$ are selected together in $S_i$.
      Also, some constants $C_{4},C_{5}$ exist s.t.
        \begin{eqnarray}
          \Delta_1 &\equiv& \max_{i=1,\ldots,N_I} \max_{k \neq l=1,\ldots,N_i} \left| \pi_{kl|i}-\pi_{k|i} \pi_{l|i} \right|  \leq  C_4 N_0^{-2} n_0, \label{eq13:sec3}\\
          \Delta_2 &\equiv& \max_{i=1,\ldots,N_I} \max_{k \neq l \neq k' \neq l'=1,\ldots,N_i} \left| \pi_{klk'l'|i}-\pi_{k|i} \pi_{l|i} \pi_{k'|i} \pi_{l'|i}\right|  \leq  C_5  N_0^{-4} n_0^3, \nonumber 
        \end{eqnarray}
      with $\pi_{klk'l'|i}$ the conditional probability that the SSUs $k,l,k',l'$ are selected together in $S_i$.
      \item[SS3:] Some constant $c_{2}>0$ exists s.t. for any PSU $i$ and for any $k \ne l$ inside:
        \begin{eqnarray} \label{eq14b:sec3}
          c_{2} N_0^{-2} n_0^{2} & \leq \pi_{kl|i}. &
        \end{eqnarray}
    \end{itemize}

\noindent It is assumed in (SS0) that the sizes $N_i$ of the PSUs are comparable, and that the numbers $n_i$ of SSUs selected inside the PSUs are also comparable. In practice, to reduce the variance associated to the first stage of sampling, the PSUs are usually grouped into strata in such a way that the PSUs inside one stratum are of similar sizes. Also, the number of selected SSUs is commonly the same for any PSU, so that all the interviewers have a comparable workload. Equations (\ref{eq8:sec3}) and (\ref{eq9:sec3}) seem therefore reasonable in practice. The assumptions (SS1)-(SS3) are similar to the assumptions (FS1)-(FS3) made for the first-stage sampling design. \\

\noindent As previously mentioned, one-stage sampling designs are a particular case of our set-up. They are obtained when $N_i=1$ for any PSU $i$ and when $n_i=1$ for any unit $i \in S_I$. In such case, assumptions (SS0)-(SS1) automatically hold while assumptions (SS2)-(SS3) vanish.

\subsection{Assumptions on the variable of interest} \label{sec23}


    \begin{itemize}
      \item[VAR1:] There exists some constants $M_1$ and $m_1>0$ such that
        \begin{eqnarray}
          & \displaystyle N^{-1} \sum_{i=1}^{N_I} \sum_{k =1}^{N_i} y_{ik}^4 & \leq M_1, \label{eq16:sec3} \\
          m_1 \leq & \displaystyle N^{-1} \sum_{i=1}^{N_I} \sum_{k =1}^{N_i} y_{ik}. & \label{eq17:sec3}
        \end{eqnarray}
      \item[VAR2:] There exists some constant $m_2>0$ such that
        \begin{eqnarray} \label{eq18:sec3}
          m_2 & \leq & N^{-2} n_I \left\{V_1(\hat{Y}_{\pi})\right\}.
        \end{eqnarray}
    \end{itemize}

\noindent It is assumed in (VAR1) that the variable of interest has a bounded moment of order four, and a mean bounded away from $0$. It is assumed in (VAR2) that the first-stage sampling variance is non-vanishing. These assumptions are fairly weak, although we may find situations under which they are not respected. The condition (\ref{eq16:sec3}) is not fulfilled for heavily skewed populations, where a non-negligible part of the individuals exhibit particularly large values for the variable of interest. This may be the case in wealth surveys, for example. Equations (\ref{eq17:sec3}) and (\ref{eq18:sec3}) are not fulfilled when we are interested in domain estimation, and when the domain size $N_d$ is negligible as compared to the population size.

\section{Consistency of estimators} \label{sec4}

\noindent We begin with determining the orders of magnitude of the components of the variance decomposition in (\ref{eq4:sec2}). The proof of Proposition \ref{prop1} follows from some moment inequalities, which are given in Section 1 
of the Supplementary Material.

\begin{prop} \label{prop1}
  Suppose that assumptions (FS1)-(FS2),(SS0)-(SS2) and (VAR1) hold. Then
    \begin{eqnarray}
      V_1(\hat{Y}_{\pi}) & = & O\left(N^2 n_I^{-1}\right), \nonumber \\
      V_2(\hat{Y}_{\pi}) & = & O\left(N^2 n_I^{-1} n_0^{-1}\right), \label{eq1:sec4} \\
      V_3(\hat{Y}_{\pi}) & = & O\left(N^2 N_I^{-1} n_0^{-1}\right). \nonumber
    \end{eqnarray}
\end{prop}

\noindent When $n_0 \to \infty$, the first variance component is the leading term and the two last ones are negligible. When $n_0$ is bounded, the first and second component have the same order of magnitude. The third component is negligible if $N_I^{-1} n_I \to 0$, and has the same order of magnitude otherwise. In practice, the third term is expected to be small as compared to the two first ones. \\

\noindent The consistency of the HT estimator is established in Proposition \ref{prop2}. The proof follows from Proposition \ref{prop1}, and is therefore omitted.

\begin{prop} \label{prop2}
  Suppose that assumptions (FS1)-(FS2),(SS0)-(SS2) and (VAR1) hold. Then the HT estimator is design-unbiased. Also, we have
    \begin{eqnarray}
      E\left[ N^{-1} \left\{\hat{Y}_{\pi}-Y\right\}\right]^2 = O(n_I^{-1})
      & \textrm{ and } & \frac{\hat{Y}_{\pi}}{Y} \longrightarrow_{Pr} 1, \label{eq2:sec4}
    \end{eqnarray}
  where $\rightarrow_{Pr}$ stands for the convergence in probability.
\end{prop}

\noindent Proposition \ref{prop2} implies that the HT estimator is $\sqrt{n}$-consistent for the true total. Also, it is important to note that the consistency of the HT-estimator requires that the sampled number of PSUs $n_I$ tends to infinity, while the consistency is not related to the behaviour of $n_0$. For example, suppose that a sample of same size $n_i=n_0$ is selected inside any PSU, so that the total number of SSUs selected is $n=n_I n_0$. Then, even if $n \to \infty$, the HT-estimator may be inconsistent if $n_I$ is bounded. In practice, it is therefore important that a large number of PSUs is selected at the first stage. \\

\noindent The consistency of the HT-estimator is proved in \cite{bre:ops:08} under the alternative assumptions:
\begin{itemize}
  \item[D4:] For any population $U$, $\min_{k \in U} \pi_k \geq \pi^* >0$ where $N \pi^* \to \infty$, and there exists $\kappa \geq 0$ such that
    \begin{eqnarray*}
    N^{0.5+\kappa} (\pi^*)^2 \to \infty & \textrm{ and } & \max_{k \in U} \sum_{l \in U;l \ne k} (\Delta_{kl})^2 = O(N^{-2\kappa}),
    \end{eqnarray*}
   where $\Delta_{kl}=\pi_{kl}-\pi_k \pi_l$.
  \item[D5:] The variable of interest satisfies
  \begin{eqnarray*}
  	\limsup \frac{1}{N} \sum_{i=1}^{N_I} \sum_{k=1}^{N_i} y_{ik}^2 \leq \infty.
  \end{eqnarray*}
\end{itemize}
Under (D4) and (D5), we have $E\left[ N^{-1} \left\{\hat{Y}_{\pi}-Y\right\}\right]^2 = o(1)$ \citep[][Lemma A.1]{bre:ops:08}. Clearly, our condition in (VAR1) on the fourth moment implies (D5). Also, it can be shown that if $n_0$ is bounded, our assumptions (FS1)-(FS2) and (SS0)-(SS2) imply (D4) with $\kappa=0$. Our stronger conditions are needed in particular to get the consistency of variance estimators, see Sections \ref{sec41} and \ref{sec42}.

\subsection{Unbiased variance estimators} \label{sec41}

\noindent We first consider the so-called Horvitz-Thompson variance estimator
    \begin{eqnarray}
      \hat{V}_{HT}(\hat{Y}_{\pi}) & = & \sum_{i,j \in S_I} \frac{\Delta_{Iij}}{\pi_{Iij}} \frac{\hat{Y}_i}{\pi_{Ii}} \frac{\hat{Y}_j}{\pi_{Ij}} + \sum_{i \in S_I} \frac{\hat{V}_{HT,i}}{\pi_{Ii}} \nonumber \\
     & = & \hat{V}_{HT,A}(\hat{Y}_{\pi})+\hat{V}_{HT,B}(\hat{Y}_{\pi}), \label{eq4:sec4}
    \end{eqnarray}
where
    \begin{eqnarray}
      \hat{V}_{HT,i} & = & \sum_{k,l \in S_i} \frac{\Delta_{kl|i}}{\pi_{kl|i}} \frac{y_{ik}}{\pi_{k|i}} \frac{y_{il}}{\pi_{l|i}}. \label{eq5:sec4}
    \end{eqnarray}

\begin{prop} \label{prop3}
  If assumptions (FS1)-(FS3),(SS0)-(SS2) and (VAR1) hold, we have:
    \begin{eqnarray}
      E\left[ N^{-2} n_I \left\{\hat{V}_{HT,A}(\hat{Y}_{\pi})-V_1(\hat{Y}_{\pi})-V_2(\hat{Y}_{\pi})\right\}\right]^2 & = & O(n_I^{-1}). \label{eq8:sec4}
    \end{eqnarray}
  If assumptions (FS1)-(FS2),(SS0)-(SS3) and (VAR1) hold, we have:
    \begin{eqnarray}
      E\left[ N^{-2} N_I n_0 \left\{\hat{V}_{HT,B}(\hat{Y}_{\pi})-V_3(\hat{Y}_{\pi})\right\}\right]^2 & = & O(n_I^{-1}). \label{eq9:sec4}
    \end{eqnarray}
  If assumptions (FS1)-(FS3),(SS0)-(SS3),(VAR1)-(VAR2) hold, we have:
    \begin{eqnarray}
      \qquad E\left[ N^{-2} n_I \left\{\hat{V}_{HT}(\hat{Y}_{\pi})-V(\hat{Y}_{\pi})\right\}\right]^2 = O(n_I^{-1}) \textrm{ and }
      \frac{\hat{V}_{HT}(\hat{Y}_{\pi})}{V(\hat{Y}_{\pi})} \rightarrow_{Pr} 1. \label{eq10:sec4}
    \end{eqnarray}
\end{prop}

\noindent The proof of Proposition \ref{prop3} is tedious but standard, and is therefore omitted. It implies that $\hat{V}_{HT}(\hat{Y}_{\pi})$ is a term by term unbiased and $\sqrt{n}$-consistent variance estimator, in the sense that $\hat{V}_{HT,A}(\hat{Y}_{\pi})$ is unbiased and $\sqrt{n}$-consistent for $V_1(\hat{Y}_{\pi})+V_2(\hat{Y}_{\pi})$, and $\hat{V}_{HT,B}(\hat{Y}_{\pi})$ is unbiased and $\sqrt{n}$-consistent for $V_3(\hat{Y}_{\pi})$. In their Theorem 3, \cite{bre:ops:00} state a similar result in case of one-stage sampling designs, for a more general class of estimators that they call local polynomial estimators. In the literature, the consistency of the HT-variance estimator is often stated as an assumption; e.g.,  
\cite{kim:par:lee:17} for two-stage sampling designs. \\

\noindent If the sampling designs used at both stages are of fixed size, we may alternatively use the Yates-Grundy variance estimator
    \begin{eqnarray}
      \hat{V}_{YG}(\hat{Y}_{\pi}) & = & -\frac{1}{2} \sum_{i \neq j \in S_I} \frac{\Delta_{Iij}}{\pi_{Iij}} \left(\frac{\hat{Y}_i}{\pi_{Ii}} - \frac{\hat{Y}_j}{\pi_{Ij}}\right)^2 + \sum_{i \in S_I} \frac{\hat{V}_{YG,i}}{\pi_{Ii}} \nonumber \\
     & = & \hat{V}_{YG,A}(\hat{Y}_{\pi})+\hat{V}_{YG,B}(\hat{Y}_{\pi}), \label{eq6:sec4}
    \end{eqnarray}
with
    \begin{eqnarray}
      \hat{V}_{YG,i} & = & -\frac{1}{2} \sum_{k \neq l \in S_i} \frac{\Delta_{kl|i}}{\pi_{kl|i}} \left(\frac{y_{ik}}{\pi_{k|i}} - \frac{y_{il}}{\pi_{l|i}} \right)^2. \label{eq7:sec4}
    \end{eqnarray}
We prove in Proposition \ref{prop4} that $\hat{V}_{YG}(\hat{Y}_{\pi})$ is also a term by term unbiased and $\sqrt{n}$-consistent variance estimator. The proof is similar to that of Proposition \ref{prop3}.

\begin{prop} \label{prop4}
  If assumptions (FS1)-(FS3),(SS0)-(SS2) and (VAR1) hold, we have:
    \begin{eqnarray}
      E\left[ N^{-2} n_I \left\{\hat{V}_{YG,A}(\hat{Y}_{\pi})-V_1(\hat{Y}_{\pi})-V_2(\hat{Y}_{\pi})\right\}\right]^2 & = & O(n_I^{-1}). \label{eq11:sec4}
    \end{eqnarray}
  If assumptions (FS1)-(FS2),(SS0)-(SS3) and (VAR1) hold, we have:
    \begin{eqnarray}
      E\left[ N^{-2} N_I n_0 \left\{\hat{V}_{YG,B}(\hat{Y}_{\pi})-V_3(\hat{Y}_{\pi})\right\}\right]^2 & = & O(n_I^{-1}). \label{eq12:sec4}
    \end{eqnarray}
  If assumptions (FS1)-(FS3),(SS0)-(SS3), (VAR1)-(VAR2) hold, we have:
    \begin{eqnarray}
      ~\qquad E\left[ N^{-2} n_I \left\{\hat{V}_{YG}(\hat{Y}_{\pi})-V(\hat{Y}_{\pi})\right\}\right]^2 = O(n_I^{-1}) \textrm{ and }
      \frac{\hat{V}_{YG}(\hat{Y}_{\pi})}{V(\hat{Y}_{\pi})} \rightarrow_{Pr} 1. \label{eq13:sec4}
    \end{eqnarray}
\end{prop}

\subsection{Simplified one-term variance estimators} \label{sec42}

\noindent Both the variance estimators $\hat{V}_{HT}(\hat{Y}_{\pi})$ and $\hat{V}_{YG}(\hat{Y}_{\pi})$ may be cumbersome in practice, since they require an unbiased and consistent variance estimator $\hat{V}_{HT,i}$ or $\hat{V}_{YG,i}$ inside any of the selected PSUs. Consider the example of self-weighted two-stage sampling designs, which are common in practice. They consist in selecting a sample of PSUs, with probabilities $\pi_{Ii}$ proportional to the size of the PSUs, and a sample of $n_0$ SSUs inside any of the selected PSUs. This leads to equal sampling weights for all the SSUs in the population, hence the name. In case of self-weighted two-stage sampling designs, systematic sampling is frequently used at the second stage. In such case, the assumption (SS3) is usually not respected. \\

\noindent A simplified variance estimator can be obtained by using $\hat{V}_{HT,A}(\hat{Y}_{\pi})$ only, or for a fixed-size sampling design $\hat{V}_{YG,A}(\hat{Y}_{\pi})$ only, see for instance \citet{sar:swe:wre:92}. Proposition \ref{prop5} states that these simplified estimators are consistent when
    \begin{eqnarray}
      \frac{V_3(\hat{Y}_{\pi})}{V_1(\hat{Y}_{\pi})+V_2(\hat{Y}_{\pi})} & \to & 0, \label{eq14:sec4}
    \end{eqnarray}
i.e. when the third component of the variance in the decomposition (\ref{eq4:sec2}) is negligible. Note that in Proposition \ref{prop5} we do not need the assumption (SS3) which guarantees a lower bound for the second-order inclusion probabilities at the second stage.

\begin{prop} \label{prop5}
  Suppose that assumptions (FS1)-(FS3), (SS0)-(SS2), (VAR1)-(VAR2) hold. Suppose that equation (\ref{eq14:sec4}) holds. Then
    \begin{eqnarray}
      E\left[ N^{-2} n_I \left\{\hat{V}_{HT,A}(\hat{Y}_{\pi})-V(\hat{Y}_{\pi})\right\}\right]^2 & = & o(1), \label{eq15:sec4} \\
      \frac{\hat{V}_{HT,A}(\hat{Y}_{\pi})}{V(\hat{Y}_{\pi})} & \longrightarrow_{Pr} & 1. \label{eq15b:sec4}
    \end{eqnarray}
  If in addition the first-stage sampling design is of fixed-size, we have
    \begin{eqnarray}
      E\left[ N^{-2} n_I \left\{\hat{V}_{YG,A}(\hat{Y}_{\pi})-V(\hat{Y}_{\pi})\right\}\right]^2 & = & o(1), \label{eq16:sec4} \\
      \frac{\hat{V}_{YG,A}(\hat{Y}_{\pi})}{V(\hat{Y}_{\pi})} & \longrightarrow_{Pr} & 1. \label{eq17:sec4}
    \end{eqnarray}
\end{prop}

\noindent The proof is immediate from Propositions \ref{prop3} and \ref{prop4}, and by using equation (\ref{eq14:sec4}). The simplified variance estimators $\hat{V}_{HT,A}(\hat{Y}_{\pi})$ and $\hat{V}_{YG,A}(\hat{Y}_{\pi})$ are simpler to compute, since they do not involve variance estimators $\hat{V}_i$ inside PSUs, but only unbiased estimators $\hat{Y}_i$ for the sub-totals over the PSUs. \\

\noindent Under the assumptions (FS1)-(FS3), (SS0)-(SS2) and (VAR1)-(VAR2), a sufficient condition for equation (\ref{eq14:sec4}) to hold is that $N_I^{-1} n_I \to 0$ (negligible first-stage sampling rate). In practice, we expect the term $V_3(\hat{Y}_{\pi})$ to have a small contribution in the overall variance even if the first-stage sampling rate is not negligible. This is illustrated in Section \ref{sec6} through a simulation study, and in Section \ref{appli:ppv} in the application to the panel for urban policy.  The two simplified variance estimators $\hat{V}_{HT,A}(\hat{Y}_{\pi})$ and $\hat{V}_{YG,A}(\hat{Y}_{\pi})$ may therefore be reasonable choices for variance estimation in practice.

\section{Case of large entropy sampling designs} \label{sec5}

\noindent In this Section, we focus on the situation when large entropy sampling designs are used at the first stage. We consider a H\'ajek-type variance estimator, and prove its consistency with limited assumptions, namely by dropping the conditions (FS2) and (FS3). Building on the work of \cite{ohl:89}, we also prove that the HT-estimator is asymptotically normally distributed. The rejective sampling design \citep{haj:64} is first considered in Section \ref{sec51}. The results are extended in Section \ref{sec52} to a class of large entropy sampling designs by using a coupling algorithm. The properties of a simplified variance estimator are studied in Section \ref{sec53}.

\subsection{Rejective sampling} \label{sec51}

\noindent The rejective (or conditional Poisson) sampling design was introduced by \cite{haj:64}. Rejective sampling in $U_I$ consists in repeatedly selecting samples by means of Poisson sampling, until the sample has the required size $n_I$. The inclusion probabilities of the Poisson sampling design are chosen so that the required inclusion probabilities $\pi_{Ii},~i \in U_I$ are respected; see for example \cite{dup:75}. The rejective sampling design has been extensively studied in the literature, see \cite{til:06} for a review. Under a rejective sampling design at the first-stage, the assumption (FS2) is implied by the assumption (FS1), see our discussion in Section \ref{sec21}. \\

\noindent We note $p_r(\cdot)$ the rejective sampling design with inclusion probabilities $\pi_{Ii}$ in the population $U_I$. Also, we note $S_{rI}$ a first-stage sample selected by means of $p_r$, and
    \begin{eqnarray}
      \hat{Y}_{r\pi} & = & \sum_{i \in S_{rI}} \frac{\hat{Y}_i}{\pi_{Ii}}
    \end{eqnarray}
the associated HT-estimator. Making use of a uniform approximation of the second-order inclusion probabilities, \cite{haj:64} proposed a very simple variance estimator for which these second-order inclusion probabilities are not needed. In our two-stage sampling context, this leads to replacing in (\ref{eq4:sec4}) the term $\hat{V}_{HT,A}(\hat{Y}_{r\pi})$ with
    \begin{eqnarray}
      \qquad \hat{V}_{HAJ,A}(\hat{Y}_{r\pi}) & = &
       \left\{\begin{array}{ll}
          \displaystyle \sum_{i \in S_{rI}} (1-\pi_{Ii}) \left(\frac{\hat{Y}_i}{\pi_{Ii}} - \hat{\hat{R}}_{r\pi} \right)^2 & \textrm{if } \hat{d}_{rI} \geq \frac{c_{I0}}{2} n_I,  \\
          0 & \textrm{otherwise},
          \end{array} \right.
       \label{eq1:sec5}
    \end{eqnarray}
with
    \begin{eqnarray}
      \hat{\hat{R}}_{r\pi} = \hat{d}_{rI}^{-1} \sum_{i \in S_{rI}} (1-\pi_{Ii}) \frac{\hat{Y}_i}{\pi_{Ii}} & \textrm{ and } & \hat{d}_{rI} = \sum_{i \in S_{rI}} (1-\pi_{Ii}), \label{eq2:sec5}
    \end{eqnarray}
and where $c_{I0}$ is defined in Lemma \ref{lem6b} (see the Supplementary Material). This leads to the global variance estimator
    \begin{eqnarray}
      \hat{V}_{HAJ}(\hat{Y}_{r\pi}) & = & \hat{V}_{HAJ,A}(\hat{Y}_{r\pi})+\hat{V}_{HT,B}(\hat{Y}_{r\pi}), \label{eq3:sec5}
    \end{eqnarray}
where $\hat{V}_{HT,B}(\hat{Y}_{r\pi})$ is defined in equation (\ref{eq4:sec4}). If the second-stage sampling designs are all of fixed-size, we could alternatively replace $\hat{V}_{HT,B}(\hat{Y}_{r\pi})$ with $\hat{V}_{YG,B}(\hat{Y}_{r\pi})$ given in equation (\ref{eq6:sec4}). \\

\noindent Note that the variance estimator $\hat{V}_{HAJ}(\hat{Y}_{r\pi})$ is truncated to avoid extreme values for $\hat{\hat{R}}_{r\pi}$. This is needed to establish its consistency, which is done in Proposition \ref{prop6}. An advantage of this variance estimator is that the first-stage second-order inclusion probabilities are not required. In particular, the condition (FS3) is not needed to prove the consistency. We also prove in Proposition \ref{prop6} that the HT-estimator is asymptotically normally distributed, by using Theorem 2.1 in \cite{ohl:89}.

\begin{prop} \label{prop6}
  Suppose that a rejective sampling design is used at the first stage. Suppose that assumptions (FS1), (SS0)-(SS2) and (VAR1) hold. Then
    \begin{eqnarray} \label{eq4:sec5}
      E\left[ N^{-2} n_I \left\{\hat{V}_{HAJ,A}(\hat{Y}_{r\pi})-V_1(\hat{Y}_{r\pi})-V_2(\hat{Y}_{r\pi})\right\}\right]^2 & = & o(1).
    \end{eqnarray}
  If in addition the assumption (VAR2) holds, then
    \begin{eqnarray} \label{eq4b:sec5}
      \frac{\hat{Y}_{r\pi}-Y}{\sqrt{V(\hat{Y}_{r\pi})}} & \longrightarrow_{\mathcal{L}} & \mathcal{N}(0,1),
    \end{eqnarray}
  where $\rightarrow_{\mathcal{L}}$ stands for the convergence in distribution. If in addition the assumption (SS3) holds, then
    \begin{eqnarray}
      \qquad E\left[ N^{-2} n_I \left\{\hat{V}_{HAJ}(\hat{Y}_{r\pi})-V(\hat{Y}_{r\pi})\right\}\right]^2 = o(1)
      \textrm{ and } \frac{\hat{V}_{HAJ}(\hat{Y}_{r\pi})}{V(\hat{Y}_{r\pi})} \rightarrow_{Pr} 1. \label{eq5:sec5}
    \end{eqnarray}
\end{prop}

\noindent The proof is given in Section 2 
of the Supplementary Material. The asymptotic normality of the HT-estimator has been proved by \cite{haj:64} for a single stage rejective sampling design, but the consistency of the H\'ajek-type variance estimator has not been rigorously established previously. Proposition \ref{prop6} has therefore its own interest, even for one-stage sampling designs. It follows that under rejective sampling at the first-stage, an approximate two-sided $100(1-2\alpha) \% $ confidence interval for $Y$ is obtained as
    \begin{eqnarray} \label{eq6:sec5}
    \left[\hat{Y}_{r\pi} \pm u_{1-\alpha} \{\hat{V}_{HAJ}(\hat{Y}_{r\pi})\}^{0.5}\right],
    \end{eqnarray}
with $u_{1-\alpha}$ the quantile of order $1-\alpha$ of the standard normal distribution.

\subsection{Other sampling designs} \label{sec52}

\noindent We consider a more general class of sampling designs at the first-stage, which are close to the rejective sampling design with respect to the Chi-square distance.
Other distance functions have been considered in the literature, such as the Hellinger distance \citep{con:14} or the total variation distance \citep{ber:cha:cle:17}.
We note $p(\cdot)$ for a fixed-size sampling design with inclusion probabilities $\pi_{Ii}$ in the population $U_I$. It is said to be close to the rejective sampling design $p_r(\cdot)$ with respect to the Chi-square distance if
    \begin{eqnarray} \label{eq1:sec52}
      \qquad d_2(p,p_r) \rightarrow 0 & \textrm{where} & d_2(p,p_r)=\sum_{s_I \subset U_I;~p_r(s_I)>0} \frac{\left\{p(s_I)-p_r(s_I)\right\}^2}{p_r(s_I)}.
    \end{eqnarray}
Equation (\ref{eq1:sec52}) holds for the Rao-Sampford \citep{sam:67} sampling design, for example.
We note $S_{pI}$ a first-stage sample selected by means of $p(\cdot)$, and the associated HT-estimator is
    \begin{eqnarray}
      \hat{Y}_{p\pi} & = & \sum_{i \in S_{pI}} \frac{\hat{Y}_i}{\pi_{Ii}}.
    \end{eqnarray}

\noindent We introduce in Algorithm \ref{algo:1} a coupling procedure to obtain the estimators $\hat{Y}_{p\pi}$ and $\hat{Y}_{r\pi}$ jointly, which is the main tool in extending the results in Proposition \ref{prop6} to $\hat{Y}_{p\pi}$. We note
    \begin{eqnarray}
      \qquad \alpha=1-d_{TV}(p,p_r) & \textrm{where} & d_{TV}(p,p_r) = \frac{1}{2} \sum_{s_I \in U_I} |p(s_I)-p_r(s_I)|
    \end{eqnarray}
is the total variation distance between $p(\cdot)$ and $p_r(\cdot)$. By using Lemma 11 in Section 3 of the Supplementary Material, it can be proved that the coupling procedure in Algorithm \ref{algo:1} leads to estimators $\hat{Y}_{r\pi}$ and $\hat{Y}_{p\pi}$ associated to the required two-stage sampling designs; see also \cite{van:16}, Theorem 2.9.

\begin{algorithm}[t]
\begin{enumerate}
	\item Draw $u$ from a uniform distribution $U[0,1]$.
    \item If $u \leq \alpha$, then:
        \begin{enumerate}
          \item Select a sample $s_I$ with probabilities $\displaystyle \frac{p(s_I) \wedge p_r(s_I)}{\alpha}$, and take $S_{rI}=S_{pI}=s_I$.
          \item For any $i \in S_{rI}=S_{pI}$, select the same second-stage sample $S_{i}$ for both $\hat{Y}_{r\pi}$ and $\hat{Y}_{p\pi}$.
        \end{enumerate}
    \item If $u > \alpha$, then:
        \begin{enumerate}
          \item Select the sample $S_{pI}$ with probabilities $\displaystyle \frac{p(s_{I}) - p_r(s_{I})}{1-\alpha}$ in the set $\{s_I \in U_I;~p(s_I)>p_r(s_I)\}$. For any $i \in S_{pI}$, select a second-stage sample $S_{i}$ for $\hat{Y}_{p\pi}$.
          \item Independently of $S_{pI}$ and of the associated second-stage samples $S_{i}$'s, select the sample $S_{rI}$ with probabilities $\displaystyle \frac{p_r(s_{I}) - p(s_{I})}{1-\alpha}$ in the set $\{s_I;~p(s_I) \leq p_r(s_I)\}$. For any $i \in S_{rI}$, select a second-stage sample $S_{i}$ for $\hat{Y}_{r\pi}$.
        \end{enumerate}
 \end{enumerate}
\caption{A coupling procedure between two-stage sampling designs} \label{algo:1}
\end{algorithm}

\begin{prop} \label{prop7a}
Suppose that the samples $S_{rI}$ and $S_{pI}$ are selected by means of the coupling procedure in Algorithm \ref{algo:1}. Then:
    \begin{eqnarray*} 
      E \left(\hat{Y}_{p\pi}-\hat{Y}_{r\pi}\right)^2 & \leq & \sum_{s_I \in U_I} |p(s_I)-p_r(s_I)| \left\{\left(\sum_{i \in s_I} \frac{Y_i}{\pi_{Ii}}-Y \right)^2 + \sum_{i \in s_I} \frac{V_i}{\pi_{Ii}^2} \right\}.
    \end{eqnarray*}
\end{prop}

\begin{prop} \label{prop7}
  Suppose that the samples $S_{rI}$ and $S_{pI}$ are selected by means of the coupling procedure in Algorithm \ref{algo:1}. Suppose that assumptions (FS1), (SS0)-(SS2) and (VAR1) hold. Suppose that $d_2(p,p_r) \to 0$. Then
    \begin{eqnarray} \label{eq1:prop7}
      E \left(\hat{Y}_{p\pi}-\hat{Y}_{r\pi}\right)^2 & = & o\left(N^2 n_I^{-1}\right).
    \end{eqnarray}
  If in addition the assumption (VAR2) holds, then
    \begin{eqnarray} \label{eq2:prop7}
      \frac{V\left(\hat{Y}_{p\pi}\right)}{V\left(\hat{Y}_{r\pi}\right)} & \rightarrow & 1.
    \end{eqnarray}
\end{prop}

\noindent The proofs of Propositions \ref{prop7a} and \ref{prop7} are given in Sections 3.2 and 3.3 
of the Supplementary Material. These propositions state that if the sampling designs $p(\cdot)$ and $p_r(\cdot)$ are close with respect to the Chi-square distance, then $E \left(\hat{Y}_{p\pi}-\hat{Y}_{r\pi}\right)^2$ is smaller than the rate of convergence of $\hat{Y}_{r\pi}$. Consequently, the results in Proposition \ref{prop6} can be extended to the sampling design $p(\cdot)$, as stated in Proposition \ref{prop8}. Similar coupling arguments are used by \cite{cha:15} to obtain asymptotic results for multistage sampling designs with stratified simple random without replacement sampling at the first stage.

\begin{prop} \label{prop8}
  Suppose that assumptions (FS1), (SS0)-(SS2), (VAR1)-(VAR2) hold, and that $d_2(p,p_r) \to 0$. Then
    \begin{eqnarray} \label{eq1:prop8}
      \frac{\hat{Y}_{p\pi}-Y}{\sqrt{V(\hat{Y}_{p\pi})}} & \longrightarrow_{\mathcal{L}} & \mathcal{N}(0,1).
    \end{eqnarray}
  If in addition the assumption (SS3) holds, we have
    \begin{eqnarray} \label{eq2:prop8}
      \qquad E\left[ N^{-2} n_I \left|\hat{V}_{HAJ}(\hat{Y}_{p\pi})-V(\hat{Y}_{p\pi}) \right| \right] = o(1) & \textrm{and} &
      \frac{\hat{V}_{HAJ}(\hat{Y}_{p\pi})}{V(\hat{Y}_{p\pi})} \rightarrow_{Pr} 1.
    \end{eqnarray}
\end{prop}

\noindent The proof is given in Section 3.4 
of the Supplementary Material. From Proposition \ref{prop8}, the two-sided $100(1-2\alpha) \% $ confidence interval given in (\ref{eq6:sec5}) is also asymptotically valid for $\hat{Y}_{p\pi}$. \\

\noindent We now turn back to the choice of the distance function. Let $X(s_I)$ denote some function of a sample $s_I$. Equation (\ref{eq1:prop7}) in Proposition \ref{prop7} is based on the inequality
    \begin{eqnarray} \label{eqa:sec52}
      \sum_{s_I \subset U_I} |p(s_I)-p_r(s_I)| X(s_I) & \leq & \sqrt{d_2(p,p_r)} \times \sqrt{\sum_{s_I \subset U_I} p_r(s_I) X(s_I)^2} \nonumber \\
                                                  & \leq & \sqrt{d_2(p,p_r)} \times \sqrt{E\{X(S_{rI})^2\}}.
    \end{eqnarray}
From equation (\ref{eqa:sec52}) and Proposition \ref{prop7a}, $X(S_{rI})$ and $X(S_{pI})$ are asymptotically equivalent if (a) $d_2(p,p_r) \to 0$, and if (b) we can control the second moment of $X(S_{rI})$. This last point may be obtained through standard algebra for rejective sampling, see Lemma 8 for example. \\

\noindent If we rather resort to the Kullback-Leibler divergence
    \begin{eqnarray}
      d_{KL}(p,p_r) & = & \sum_{s_I \subset U_I;~p_r(s_I)>0} p(s_I) \log\left\{\frac{p(s_I)}{p_r(s_I)}\right\},
    \end{eqnarray}
we can obtain the similar inequality
    \begin{eqnarray*}
      \sum_{s_I \subset U_I} |p(s_I)-p_r(s_I)| X(s_I) \leq
      \sqrt{d_{KL}(p,p_r)} \times \sqrt{\frac{4}{3} E\{X(S_{rI})^2\} + \frac{2}{3} E\{X(S_{pI})^2\}}.
    \end{eqnarray*}
Consequently, we may alternatively demonstrate that $X(S_{rI})$ and $X(S_{pI})$ are asymptotically equivalent if (a') $d_{KL}(p,p_r) \to 0$, if (b) we can control the second moment of $X(S_{rI})$, and if (c) we can control the second moment of $X(S_{pI})$. This last point is difficult to prove for a general sampling design.

\subsection{A simplified variance estimator} \label{sec53}

\noindent The variance estimator $\hat{V}_{HAJ}(\hat{Y}_{r\pi})$ proposed in (\ref{eq3:sec5}) has been proved to be consistent for large entropy sampling designs, with limited assumptions on the first-stage sampling design. However, unbiased and consistent variance estimators are required inside the PSUs, which can be cumbersome for a data user. It is stated in Proposition \ref{prop9} that the simplified one-term variance estimator $\hat{V}_{HAJ,A}(\hat{Y}_{r\pi})$ is consistent, provided that the third component of the variance in the decomposition (\ref{eq4:sec2}) is negligible. The proof readily follows from Propositions \ref{prop6} and \ref{prop8}, and is therefore omitted. Note that the assumption (SS3) providing a lower bound for the second order inclusion probabilities at the second stage is not needed any more.

\begin{prop} \label{prop9}
  Suppose that assumptions (FS1), (SS0)-(SS2), (VAR1)-(VAR2) hold. Suppose that equation (\ref{eq14:sec4}) holds. If a rejective sampling design $p_r$ is used at the first-stage, we have
    \begin{eqnarray}
      E\left[ N^{-2} n_I \left\{\hat{V}_{HAJ,A}(\hat{Y}_{r\pi})-V(\hat{Y}_{r\pi})\right\}\right]^2 & = & o(1), \label{eq1:prop9} \\
      \frac{\hat{V}_{HAJ,A}(\hat{Y}_{r\pi})}{V(\hat{Y}_{r\pi})} & \longrightarrow_{Pr} & 1. \label{eq2:prop9}
    \end{eqnarray}
  If the first-stage sampling design $p$ is such that $d_2(p,p_r) \to 0$, then
    \begin{eqnarray}
      E\left[ N^{-2} n_I \left|\hat{V}_{HAJ,A}(\hat{Y}_{p\pi})-V(\hat{Y}_{p\pi})\right|\right] = o(1) & \textrm{and} &
      \frac{\hat{V}_{HAJ,A}(\hat{Y}_{p\pi})}{V(\hat{Y}_{p\pi})} \rightarrow_{Pr} 1. \label{eq3:prop9}
    \end{eqnarray}
\end{prop}

\section{Simulation study}\label{sec6}

\noindent A simulation study was conducted to evaluate the asymptotic properties of the H\'ajek-type variance estimators $\hat{V}_{HAJ}(\hat{Y}_{\pi})$ and $\hat{V}_{HAJ,A}(\hat{Y}_{\pi})$. Three populations $U_1,U_2,U_3$ of $N_I=2,000$ PSUs were generated. The number of SSUs per PSU were randomly generated, with mean $N_0=40$ and with a coefficient of variation equal to 0, 0.03 and 0.06 for population 1, 2, and 3 respectively. The PSUs are therefore of equal size in the first population. \\

\noindent In each population, a value $\nu_i$ was generated for any PSU $i$ from a standard normal distribution. Three variables were generated, for any SSU $k$ inside PSU $i$, in each population according to the model
    \begin{eqnarray*}
      y_{ikh} = \lambda + \sigma \nu_i + [ \rho_h^{-1} (1-\rho_h)]^{0.5} \sigma \varepsilon_k,
    \end{eqnarray*}
where $\lambda = 20$, $\sigma = 2$, where $\varepsilon_k$ was generated from a standard normal distribution, and $\rho_h$ was such that the intra-cluster correlation coefficient ($ICC$) was approximately 0.1, 0.2 and 0.3 for $h=1,2$ and 3 respectively. \\

\noindent From each population, we repeated $R=1,000$ times the following two-stage sampling design. A first-stage sample $S_I$ of $n_I=20, 40, 100$ or 200 PSUs was selected by means of a rejective sampling design, with inclusion probabilities $\pi_{Ii}$ proportional to the size $N_i$. A second-stage sample $S_i$ of $n_i=n_0=5$ or 10 was selected inside any $i \in S_I$ by simple random sampling without replacement. In each sample, we computed the HT-estimator $\hat{Y}_\pi$ and the H\'ajek-type variance estimators $\hat{V}_{HAJ,A}(\hat{Y}_\pi)$ and $\hat{V}_{HAJ}(\hat{Y}_\pi)$. \\

\noindent As a measure of bias of a variance estimator $\hat{V}$, we computed the Monte Carlo percent relative bias
    \begin{eqnarray*}
    \mathrm{RB}_{MC}(\hat{V}) = \frac{ \displaystyle \frac{1}{R} \sum_{r=1}^R \hat{V}^{(r)} - V(\hat{Y}_\pi) }{ V(\hat{Y}_\pi) }  \times 100,
    \end{eqnarray*}
with $\hat{V}^{(r)}$ the value of the estimator in the $r$th sample, and $\mathrm{V}(\hat{Y}_\pi)$ the exact variance. The Monte Carlo percent relative stability,
    \begin{eqnarray*}
    \mathrm{RS}_{MC}(\hat{V}) =
    \frac{ \displaystyle \left\{ \frac{1}{R} \sum_{r=1}^R  \left[ \hat{V}^{(r)} - \mathrm{V}(\hat{Y}_\pi) \right]^2 \right\}^{1/2} }{\mathrm{V}(\hat{Y}_\pi)} \times 100,
    \end{eqnarray*}
was calculated as a measure of variability of $\hat{V}$. We also calculated the error rates of the normality-based confidence interval given in (\ref{eq6:sec5}), with nominal one-tailed error rate of 2.5 \% in each tail.\\

\noindent The results are presented in Table \ref{tab:pop3} for the population 3. We observed no qualitative difference with populations 1 and 2, and the results are therefore omitted for conciseness. As expected, the variance estimator $\hat{V}_{HAJ}(\hat{Y}_\pi)$ is almost unbiased in any case, with $\mathrm{RB}_{MC}$ lower than 2\% in absolute value. The stability $\mathrm{RS}_{MC}$ decreases with $n_I$ but not with $n_i$, as expected. The bias of the simplified variance estimator $\hat{V}_{HAJ,A}(\hat{Y}_\pi)$ is comparable with a small first-stage sampling fraction, but increases with $n_I/N_I$. Even with the largest sampling fraction, the bias of $\hat{V}_{HAJ,A}(\hat{Y}_\pi)$ is limited and no greater than $7 \% $. This supports the fact that the term of variance $V_3(\hat{Y}_{\pi})$ in the decomposition (\ref{eq1:sec4}) has a small contribution to the global variance. Both variance estimators perform similarly in terms of stability, with $\mathrm{RS}_{MC}$ being slightly larger for $\hat{V}_{HAJ,A}(\hat{Y}_\pi)$ with the largest sampling fraction. The coverage probabilities are well respected in any case, lying between 93\% and 95\%.

\begin{table}[h!]\caption{Percent relative biases, percent relative stabilities and coverage probabilities of $\widehat{V}_{HAJ,A}(\widehat{Y}_\pi)$ and $\widehat{V}_{HAJ}(\widehat{Y}_\pi)$ in population 3} \label{tab:pop3}
	\centering
	\begin{tabular}{lllcccccc}
		\hline
		& & & \multicolumn{2}{c}{$RB_{MC}$} & \multicolumn{2}{c}{$RS_{MC}$} & \multicolumn{2}{c}{$CI_{MC}$}  \\
		\cline{4-5} \cline{6-7} \cline{8-9}
		$ICC$ & $n_I$ & $n_i$ & $\widehat{V}_{HAJ,A}(\widehat{Y}_\pi)$ & $\widehat{V}_{HAJ}(\widehat{Y}_\pi)$& $\widehat{V}_{HAJ,A}(\widehat{Y}_\pi)$ & $\widehat{V}_{HAJ}(\widehat{Y}_\pi)$ & $\widehat{V}_{HAJ,A}(\widehat{Y}_\pi)$ & $\widehat{V}_{HAJ}(\widehat{Y}_\pi)$ \\
		\hline
		0.1 & 20 & 5 & 0.08 & 0.70 & 33.58 & 33.59 & 0.94 & 0.94 \\
		&  & 10 & -0.98 & -0.57 & 31.30 & 31.30 & 0.93 & 0.93 \\
		& 40 & 5 & -1.00 & 0.24 & 21.59 & 21.56 & 0.94 & 0.94 \\
		&  & 10 & -2.66 & -1.84 & 21.85 & 21.77 & 0.93 & 0.93 \\
		& 100 & 5 & -3.23 & -0.08 & 14.02 & 13.64 & 0.94 & 0.94 \\
		&  & 10 & -2.36 & -0.27 & 14.34 & 14.15 & 0.95 & 0.95 \\
		& 200 & 5 & -6.59 & -0.19 & 11.17 & 9.03 & 0.94 & 0.94 \\
		&  & 10 & -4.15 & 0.17 & 10.42 & 9.57 & 0.94 & 0.95 \\
		\hline
		0.2 & 20 & 5 & -0.37 & 0.05 & 33.13 & 33.13 & 0.93 & 0.93 \\
		&  & 10 & -0.80 & -0.57 & 32.03 & 32.02 & 0.93 & 0.93 \\
		& 40 & 5 & -0.82 & 0.01 & 22.20 & 22.18 & 0.94 & 0.94 \\
		&  & 10 & -2.17 & -1.71 & 21.99 & 21.94 & 0.93 & 0.93 \\
		& 100 & 5 & -2.25 & -0.13 & 14.07 & 13.89 & 0.95 & 0.95 \\
		&  & 10 & -1.75 & -0.56 & 14.34 & 14.25 & 0.94 & 0.95 \\
		& 200 & 5 & -4.54 & -0.17 & 10.20 & 9.14 & 0.94 & 0.94 \\
		&  & 10 & -2.22 & 0.28 & 9.96 & 9.72 & 0.94 & 0.94 \\
		\hline
		0.3 & 20 & 5 & -0.72 & -0.43 & 32.89 & 32.88 & 0.94 & 0.94 \\
		&  & 10 & -0.69 & -0.54 & 32.39 & 32.39 & 0.93 & 0.93 \\
		& 40 & 5 & -0.77 & -0.19 & 22.58 & 22.56 & 0.94 & 0.94 \\
		&  & 10 & -1.85 & -1.55 & 22.02 & 21.99 & 0.93 & 0.93 \\
		& 100 & 5 & -1.63 & -0.14 & 14.09 & 14.00 & 0.95 & 0.95 \\
		&  & 10 & -1.44 & -0.67 & 14.29 & 14.24 & 0.95 & 0.95 \\
		& 200 & 5 & -3.26 & -0.16 & 9.80 & 9.25 & 0.95 & 0.95 \\
		&  & 10 & -1.29 & 0.32 & 9.83 & 9.75 & 0.95 & 0.95 \\
		\hline
	\end{tabular}
\end{table}

\section{Illustration on the panel for urban policy} \label{appli:ppv}

\noindent We consider an application to the Panel for Urban Policy (PUP), which is the original motivation for this work. This is a panel survey in four waves, performed by the French General Secretariat of the Inter-ministerial Committee for Cities (SGCIV) and conducted between 2011 and 2014. The scope of the survey is the collection of various information about security, employment, precariousness, schooling and health, for people living in the Sensitive Urban Zones (ZUS). The initial panel $S_I$ is selected through two-stage sampling, with districts as PSUs and households as SSUs. The individuals in the selected households are comprehensively surveyed. \\

\noindent At the first stage, the population $U_I$ of districts is partitioned into $H=4$ strata defined according to the progress of the urban renewal program. A stratified sample $S_I$ of $n_I=40$ districts is selected, with probabilities proportional to the number of main dwellings. The first-stage inclusion probabilities range from $0.04$ to $0.67$, for a first-stage sampling rate of approximately $0.09$. Inside any selected district $i$, a sample $S_i$ of $n_i$ households is selected with equal probabilities. The sample of households is prone to unit non-response, but this issue is not considered here for the sake of simplicity. In this illustration, the sample of responding households is viewed as the true sample. In summary, the data set is a sample of $1,065$ households obtained by stratified two-stage sampling. \\

\noindent We are interested in four variables related to security, town planning and residential mobility. The variable $y_1$ gives the perceived reputation of the district (good, fair, poor, no opinion). The variable $y_2$ indicates if a member of the household has witnessed trafficking (never, rarely, sometimes, no opinion). The variable $y_3$ indicates if some significant roadworks have been done in the neighborhood in the twelve last months (yes, no, no opinion). The variable $y_4$ indicates if the households intends to leave the district during the next twelve next months (certainly/probably, certainly not, probably not, no opinion). For any possible characteristic $c$ of some variable $y$, we are interested in the proportion
    \begin{eqnarray} \label{eq1:appli}
      p_{c} = \frac{\sum_{h=1}^H \sum_{i=1}^{N_{Ih}} Y_{i}}{\sum_{h=1}^H \sum_{i=1}^{N_{Ih}} N_{i}} & \textrm{ with } & Y_{i}=\sum_{k=1}^{N_i} 1(y_{ik}=c),
    \end{eqnarray}
and where $N_{Ih}$ is the number of PSUs in the stratum $h$.
The proportion $p_c$ is estimated by its substitution estimator
    \begin{eqnarray} \label{eq2:appli}
      \hat{p}_{c} = \frac{\sum_{h=1}^H \sum_{i \in S_{Ih}} \frac{\hat{Y}_i}{\pi_{Ii}}}{\hat{N}_{\pi}} & \textrm{ with } & \hat{N}_{\pi} \equiv \sum_{h=1}^H \sum_{i \in S_{Ih}} \sum_{k \in S_i} \frac{1}{\pi_{Ii} \pi_{k|i}},
    \end{eqnarray}
and where $S_{Ih}$ is the sample of PSUs in the stratum $h$.

\noindent For each proportion, we consider the two variance estimators presented in Section \ref{sec5}. We first compute the linearized variable of $p_c$, which is
    \begin{eqnarray} \label{eq3:appli}
      e_{ik} & = & \frac{1}{\hat{N}_{\pi}} \{1(y_{ik}=c)-\hat{p}_{c}\}.
    \end{eqnarray}
We then compute the variance estimator in (\ref{eq3:sec5}) by replacing the variable $y_{ik}$ with $e_{ik}$, and without truncating the first term of variance for simplicity. With stratified sampling at the first stage, and since the second-stage samples are selected with equal probabilities, this leads to the variance estimator
    \begin{eqnarray} \label{eq4:appli}
      \hat{V}_{HAJ}(\hat{p}_{c}) & = & \hat{V}_{HAJ,A}(\hat{p}_{c}) + \hat{V}_{HT,B}(\hat{p}_{c}), \\
      \textrm{with } \hat{V}_{HAJ,A}(\hat{p}_{c}) & = & \sum_{h=1}^4 \sum_{i \in S_{Ih}} (1-\pi_{Ii}) \left(\frac{\hat{E}_i}{\pi_{Ii}} - \hat{\hat{R}}_{eh\pi} \right)^2, \nonumber \\
      \textrm{with } \hat{V}_{HT,B}(\hat{p}_{c}) & = & \sum_{h=1}^4 \sum_{i \in S_{Ih}} \frac{N_i^2}{\pi_{Ii}} \left(\frac{1}{n_i}-\frac{1}{N_i}\right) s_{ei}^2, \nonumber
    \end{eqnarray}
and where
    \begin{eqnarray}  \label{eq5:appli}
      \hat{\hat{R}}_{eh\pi} = \frac{\sum_{i \in S_{Ih}} (1-\pi_{Ii}) \frac{\hat{E}_i}{\pi_{Ii}}}{\sum_{i \in S_{Ih}} (1-\pi_{Ii})} & \textrm{ with } & \hat{E}_i = \sum_{k \in S_i} \frac{e_{ik}}{\pi_{k|i}}, \\
      s_{ei}^2 = \frac{1}{n_i-1} \sum_{k \in S_i} (e_{ik}-\bar{e}_i)^2 & \textrm{ with } & \bar{e}_i = \frac{1}{n_i} \sum_{k \in S_i} e_{ik}. \nonumber
    \end{eqnarray}
The second, simplified variance estimator is $\hat{V}_{HAJ,A}(\hat{p}_{c})$, obtained from equation (\ref{eq4:appli}) by dropping the second component. \\

\noindent The two variance estimators are then plugged into a normality-based confidence interval, with a nominal one-tailed error rate of 2.5 \% . The results are presented in Table \ref{result:illust}, and show almost identical performance of both variance estimators.

\begin{table*}
  \caption{Substitution estimator of the marginal proportions and normality-based Confidence Intervals (CI) for four variables} \label{result:illust}
  \begin{tabular}{|l|cccc|} \hline
                                  & \multicolumn{4}{|c|}{Perceived Reputation of District Status} \\ \hline
                                  & Good               & Fair            & Poor              & No opinion \\
    Estimator $\hat{p}_c$         & $0.218$            & $0.227$         & $0.527$           & $0.028$ \\
    CI with $\hat{V}_{HAJ}$       & [0.182,0.253]      & [0.205,0.250]   & [0.485,0.569]     & [0.018,0.038] \\
    CI with $\hat{V}_{HAJ,A}$     & [0.183,0.252]      & [0.206,0.248]   & [0.486,0.568]     & [0.019,0.038] \\ \hline \hline
                                  & \multicolumn{4}{|c|}{Witnessed trafficking} \\ \hline
                                  & Never              & Rarely          & Sometimes         & No opinion \\
    Estimator $\hat{p}_c$         & $0.582$            & $0.053$         & $0.163$           & $0.049$ \\
    CI with $\hat{V}_{HAJ}$       & [0.537,0.628]      & [0.037,0.068]   & [0.135,0.192]     & [0.036,0.063] \\
    CI with $\hat{V}_{HAJ,A}$     & [0.538,0.627]      & [0.038,0.068]   & [0.136,0.191]     & [0.037,0.062] \\ \hline \hline
                                  & \multicolumn{4}{|c|}{Roadworks in neighborhood} \\ \hline
                                  & Yes                & No              & No opinion        & \\
    Estimator $\hat{p}_c$         & $0.463$            & $0.503$         & $0.034$           & \\
    CI with $\hat{V}_{HAJ}$       & [0.398,0.528]      & [0.434,0.572]   & [0.022,0.045]     & \\
    CI with $\hat{V}_{HAJ,A}$     & [0.399,0.527]      & [0.435,0.572]   & [0.023,0.044]     & \\ \hline \hline
                                  & \multicolumn{4}{|c|}{Intention to leave the district} \\ \hline
                                  & Certainly/Probably & Probably not    & Certainly not     & No opinion \\
    Estimator $\hat{p}_c$         & $0.275$            & $0.129$         & $0.562$           & $0.034$ \\
    CI with $\hat{V}_{HAJ}$       & [0.255,0.295]      & [0.098,0.159]   & [0.531,0.594]     & [0.025,0.043]\\
    CI with $\hat{V}_{HAJ,A}$     & [0.257,0.292]      & [0.099,0.158]   & [0.532,0.593]     & [0.036,0.042]\\ \hline
  \end{tabular}
\end{table*}

\section{Discussion} \label{sec7}

\noindent In this article, we proposed an asymptotic set-up for the study of two-stage sampling designs. We gave general conditions under which the Horvitz-Thompson estimator is consistent, and under which usual variance estimators are consistent. In case of large entropy sampling designs at the first stage, we also proved that the Horvitz-Thompson estimator is asymptotically normally distributed and that a truncated H\'ajek-like variance estimator is consistent. When the first-stage sampling fraction is negligible, simplified variance estimators are also shown to be consistent, under limited assumptions. \\

\noindent Multistage sampling designs are often used at baseline for longitudinal household surveys. If we wish to perform longitudinal estimations, individuals from the initial sample are followed over time. If we also wish to perform cross-sectional estimations at several times, additional samples are selected at further waves and mixed with the individuals originally selected. Even in the simplest case when estimations are produced at baseline with a single sample, variance estimation is challenging due to the different sources of randomness which need to be accounted for: this includes not only the sampling design, but also unit non-response, item non-response and the corresponding statistical treatments. Variance estimation in such more realistic context is an important matter for further investigation.



%
%

\end{document}